\documentclass[multphys,vecphys]{svmult}

\usepackage{makeidx}         % allows index generation
\usepackage{graphicx}        % standard LaTeX graphics tool
\usepackage{multicol}        % used for the two-column index
\usepackage[bottom]{footmisc}% places footnotes at page bottom
\def\gtwid{\mathrel{\raise.3ex\hbox{$>$\kern-.75em\lower1ex\hbox{$\sim$}}}}
\def\ltwid{\mathrel{\raise.3ex\hbox{$<$\kern-.75em\lower1ex\hbox{$\sim$}}}}

\newcommand{\ZIP}{{\small ZIP}}
\newcommand{\ZIPs}{{\small ZIP}s}
\newcommand{\TES}{{\small TES}}
\newcommand{\TESs}{{\small TES}s}
\newcommand{\CDMS}{{\small CDMS}}
\newcommand{\CDMSI}{{\small CDMS\,I}}
\newcommand{\CDMSII}{{\small CDMS\,II}}
\newcommand{\WIMP}{{\small WIMP}}

\newcommand{\etal}{et al}   			% et al in appropriate format, the following . 
\newcommand{\pl}{{\it Phys. Lett. }} 		% Physics Letters, note space at end

\def\perday{d$^{-1}$}
\def\perkg{kg$^{-1}$}
\def\iru{\perkg~\perday}
\def\perkgd{\perkg~\perday}

\makeindex             % used for the subject index
\begin{document}

\title*{The SuperCDMS Experiment}
\authorrunning{R.W.~Schnee et al.}
\author{R.W.~Schnee\inst{1}}
%\and
%Name of Author\inst{2}}
%1Case 2Brown 3Florida 4Stanford 5Fermilab 6UCSB 7Minn 8UCB 9UCDenver 
%9LBNL
\author{R.W.~Schnee\inst{1}\and D.S.~Akerib\inst{1}\and M.J.~Attisha\inst{2}\and 
        C.N.~Bailey\inst{1}\and L.~Baudis\inst{3}\and 
        D.A.~Bauer\inst{4}\and P.L.~Brink\inst{5}\and P.P.~Brusov\inst{1}\and
	R.~Bunker\inst{6}\and 
	B.~Cabrera\inst{5}\and D.O.~Caldwell\inst{6}\and
	C.L.~Chang\inst{5}\and J.~Cooley\inst{5}\and
	M.B.~Crisler\inst{4}\and P.~Cushman\inst{7}\and
	P.~Denes\inst{8}\and
	M.R.~Dragowsky\inst{1}\and L.~Duong\inst{7}\and 
	%R.~Ferril\inst{6}\and 
	J.~Filippini\inst{9}\and R.J.~Gaitskell\inst{2}\and
	S.R.~Golwala\inst{10}\and
	D.R.~Grant\inst{1}\and R.~Hennings-Yeomans\inst{1}\and 
	D.~Holmgren\inst{4}\and M.E.~Huber\inst{11}\and 
	K.~Irwin\inst{12}\and A.~Lu\inst{8}\and R.~Mahapatra\inst{6}\and
	%V.~Mandic\inst{9}\and 
	P.~Meunier\inst{9}\and N.~Mirabolfathi\inst{9}\and H.~Nelson\inst{6}\and 
	%R.~Nelson\inst{6}\and 
	R.W.~Ogburn\inst{5}\and 
	E.~Ramberg\inst{4}\and 
	%W.~Rau\inst{9}\and 
	A.~Reisetter\inst{7}\and 
	T.~Saab\inst{3}\and B.~Sadoulet\inst{9,8}\and 
	J.~Sander\inst{6}\and 
	%C.~Savage\inst{6}\and
	D.N.~Seitz\inst{9}\and 
	B.~Serfass\inst{9}\and K.M.~Sundqvist\inst{9}\and J-P.F.~Thompson\inst{2}\and 
	S.~Yellin\inst{6}\and J.~Yoo\inst{4} 
	\and B.A.~Young\inst{13}} 
\institute{Department of Physics, 
        Case Western Reserve University, Cleveland, OH 44106, USA
\texttt{schnee@case.edu}
\and Department of Physics, Brown University, Providence, RI 02912, USA
\and Department of Physics, University of Florida, Gainesville, FL 32611, USA
\and Fermi National Accelerator Laboratory, Batavia, IL 60510, USA
\and Department of Physics, Stanford University, Stanford, CA 94305, USA
\and Department of Physics, University of California, Santa Barbara, CA 93106, USA
\and School of Physics \& Astronomy, University of Minnesota, Minneapolis, MN 55455, USA
\and Lawrence Berkeley National Laboratory, Berkeley, CA 94720, USA
\and Department of Physics, University of California, Berkeley, CA 94720, USA
\and California Institute of Technology, Pasadena, CA, 91125, USA
\and Department of Physics, University of Colorado at Denver and
Health Sciences Center, Denver, CO 80217, USA
\and National Institute of Standards and Technology, Boulder, CO
80303, USA
\and Department of Physics, Santa Clara University, Santa Clara, CA 95053, USA}

\maketitle

\section{Introduction}
\label{sec:introduction}

Nonluminous, nonbaryonic, 
weakly interacting massive particles 
(WIMPs)~\cite{lee,primack} may constitute most of the matter 
in the universe~\cite{bergstrom}. WIMPs are expected to be in 
%and form 
a roughly isothermal Galactic halo. 
%These WIMPs 
They would 
interact elastically with nuclei, generating a recoil energy of a few 
tens of~keV, at a rate $\ltwid$1~event~\iru 
~\cite{primack,lewin,jkg}.

\begin{figure}[tbp]
\centering
\includegraphics[width=2.88in]{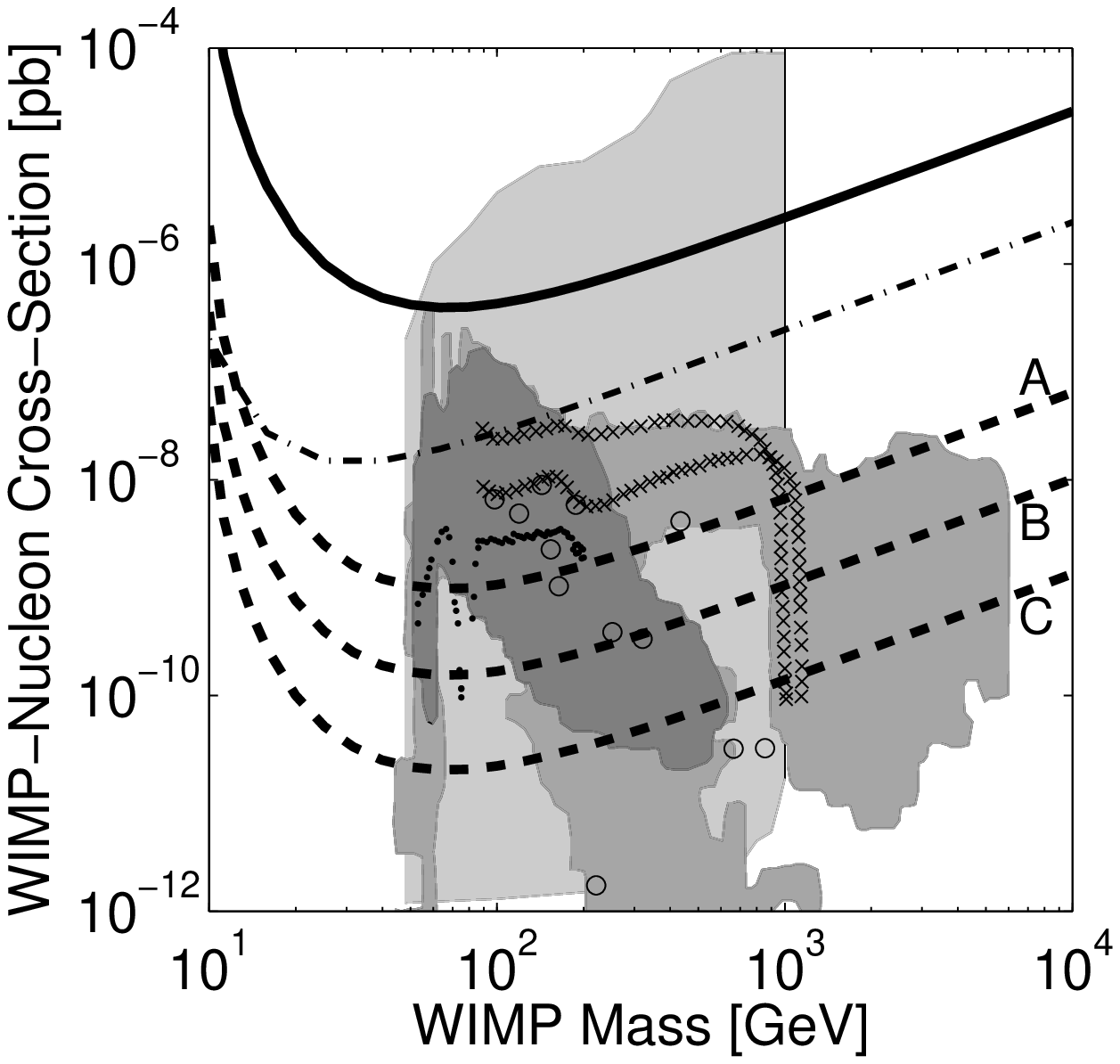}
\includegraphics[width=1.65in]{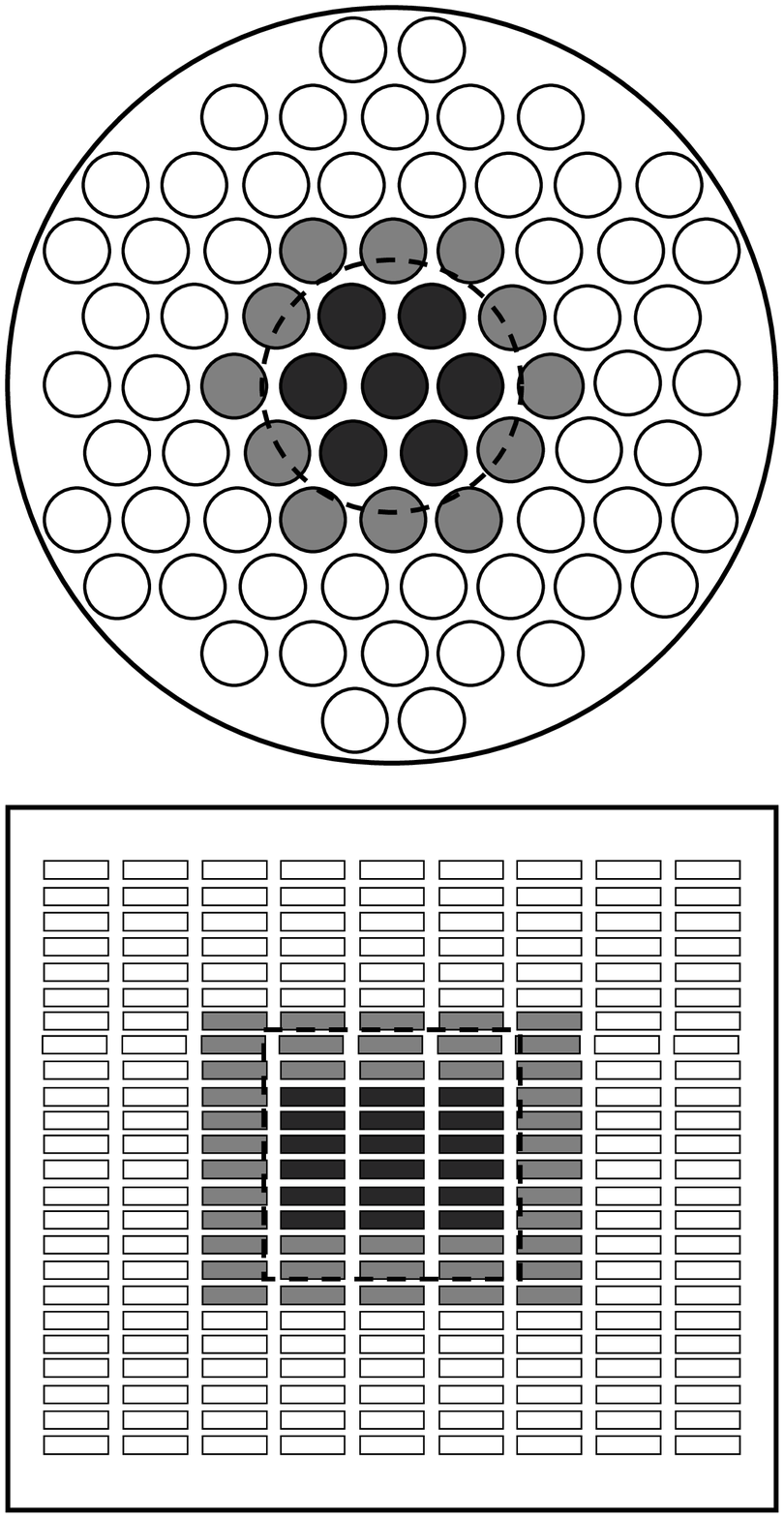}
\caption{\small {\it Left}: 
Reach of SuperCDMS phases A, B,
and C (dashed curves) with
current \CDMSII\ limit~\cite{brink} (solid curve) and sensitivity goal
(dot-dashed curve).
The lightest grey region results from a scan of MSSM parameter
space~\cite{kim}.
SuperCDMS will probe nearly all
split-supersymmetry models 
($\times$'s~\cite{Giudice04} and dots~\cite{Pierce04}) and
much of the mSUGRA region~\cite{Baltz04} (medium grey), 
including 
most post-LEP benchmark points (circles)~\cite{Battaglia03} and
nearly all the subset (dark grey) consistent with 
a supersymmetric interpretation of the
muon $g-2$ measurement.
{\it Right}: Top view (above side view) of SuperCDMS cryostat
showing 
deployment of 
7 towers of 6 detectors each in 25-kg phase A (darkest circles), 
19 towers of 12 detectors each in 150-kg phase B (grey circles), 
and 73 towers of 24 detectors each in 1-ton phase C (all circles).
The cryostat will be $3\times$ larger than the CDMS cryostat (dashes) 
in each dimension.
}
\label{fig:SuperCDMS_reach}
\end{figure}

Supersymmetry provides a natural 
\WIMP\ candidate in the form of the lightest 
superpartner~\cite{jkg}. 
Figure~\ref{fig:SuperCDMS_reach} shows the 
\WIMP\ masses and cross sections that 
are consistent with the three primary approaches to supersymmetry.
Regardless of the theoretical philosophy, 
sensitivity from elastic-scattering experiments
to \WIMP-nucleon scalar cross-sections in the range 
10$^{-46}$--10$^{-44}$\,cm$^2$
would be of great interest. %,
%particularly if the muon $g-2$ result is due to supersymmetry.

To probe to such small cross sections,
it helps
%is desirable 
to operate without backgrounds,
so that the search sensitivity 
is directly proportional to the detector mass $\times$ exposure time ($MT$).
Under subtraction of an estimated background, 
the sensitivity 
becomes proportional to
$\sqrt{MT}$. Ultimately, the subtraction becomes limited by systematics, 
preventing further improvement in sensitivity. 
%Their excellent background rejection makes using 
\CDMS\ \ZIP\ detectors~\cite{zips} have
excellent background rejection, making them 
the most proven means by which to operate an experiment without backgrounds.

\CDMS\ \ZIP\ detectors allow discrimination between \WIMP\ nuclear recoils
and background
electron recoils through two effects.  First, for a given 
energy, recoiling electrons are more ionizing than recoiling nuclei, 
resulting in a higher ratio of ionization to phonon signal, called 
``ionization yield."  Second, the athermal phonon signals due to 
nuclear recoils have longer rise times and occur later than those due 
to electron recoils.  %,
%especially electron recoils 
%within a few $\mu$m of a detector's surface~\cite{clarke,mandic}. 
For recoils within a few $\mu$m of a detector's surface (primarily 
from low-energy electrons), the charge collection is 
incomplete~\cite{shutt}, making discrimination based on ionization 
yield less effective. But these events can be effectively rejected by 
phonon timing cuts because they have, on the average, even faster 
phonon signals than those from bulk electron 
recoils~\cite{clarke,mandic}. 

In order to probe to smaller \WIMP-nucleon cross sections,
we plan to increase the detector mass of the experiment in
several phases, resulting in a ton-scale SuperCDMS experiment
(previously called CryoArray~\cite{cryoarray, schneeCryo}).
To maximize the discovery potential of the experiment, each phase 
will have an expected background smaller than one event. 
The excellent characterization of backgrounds and the 
information from ZIP detectors on each event would minimize the
ambiguity of a discovery.  
The low energy
thresholds 
and small ``quenching'' factors 
for both the ionization and the phonon measurements
allow the requirement of a 
positive signal for both energy measurements, providing immunity to
artifacts that may mimic a \WIMP\ signal.
The $\sim$keV energy resolutions and $\sim$mm position resolution
would allow detailed tests of consistency with a \WIMP\ signal.
%Although the experiment provides no directional information on the
%incident WIMP, 
A test for annual modulation would depend only on accurately knowing
the efficiency for WIMPs over time, since there would be no
additional backgrounds.

To keep backgrounds negligible during the phases of SuperCDMS,
improvements are needed
in the level and discrimination of backgrounds, described in 
Sections~\ref{sect:backgrounds} and~\ref{sect:rejection} respectively.
To achieve an exposure of 500~ton~d within a
reasonable time and budget,
manufacturability
and detector production  
rates must be improved, as discussed in
Section~\ref{sect:build}.

\section{Reduction of Backgrounds}
\label{sect:backgrounds}

%Worrisome background interactions may arise due to neutrons, photons, or
%electrons.  We discuss each in turn.

\subsection{Neutrons}
\label{sect:neutrons}

A neutron of kinetic energy $\sim$2~MeV can cause a  
recoil that is indistinguishable from 
one caused by a WIMP.
Polyethylene shielding and an active scintillator veto are used to
minimize or reject the neutron background.
For the \CDMSII\ experiment at the Soudan Mine,
%the dominant source of
most neutrons that cause unvetoed nuclear recoils in
Ge come from the
$\sim220~$GeV muons that 
penetrate to and 
interact in
the rock surrounding the \CDMSII\ experimental hall. 
Simulations indicate the rate of unvetoed
neutron-induced recoils from showers in the rock at Soudan is
$3\!\times\!10^{-4}$~\perkgd,
sufficiently low 
that it should 
%to 
be important only for exposures significantly larger
than those planned for \CDMSII\ (1000~kg~d).
To reduce this background for the 
$10\times$ to $500\times$ larger exposures
of SuperCDMS, we plan to build the experiment at
SNOLab, where the increased depth suppresses
the dominant neutron backgrounds 
by over two orders of magnitude compared to Soudan.  
%ADD COMMENTS ON SHIELDING< CLEANLINESS REQUIREMENTS

\subsection{Electron and Photon Backgrounds}
\label{sect:embackgrounds}

\CDMS\ \ZIP\ discrimination of electron-recoil events based on ionization 
yield is essentially
perfect for electrons or photons interacting in the bulk of the detector.
However, events within $\sim$35~$\mu$m of the detector surface suffer 
ionization-yield suppression, and events in the first $\sim$1~$\mu$m lose so
much ionization that they may be misidentified as nuclear recoils.
Phonon timing provides rejection of 97\% of these ``surface
events'' while keeping 70\% of true nuclear recoils.

\begin{table}[tbp]
\centering
\caption{
Mean event rates 
between 15--45~keV recoil energy
in the inner Ge detectors of \CDMSII\
and SuperCDMS, per 500 ton day exposure. 
\CDMSII\ photon ($\gamma$) and electron ($\beta$) rates 
of events with
full ionization yield (``Bulk'') and reduced ionization yield
(``Surface'')
are 
inferred from calibrations, simulations, and measurements.
%and leakage into the nuclear-recoil signal region with
%phonon-timing cuts applied
%are inferred from measurements and simulations.
Leakage into the nuclear-recoil signal region with
phonon-timing cuts applied is based on calibrations.
}
\begin{tabular}{|c|c|r|c|r|r|r|r|r|c|} 
\hline

\hline
& \multicolumn{6}{|c|}{CDMS II} & \multicolumn{3}{|c|}{SuperCDMS} \\  
    &   \multicolumn{2}{|c|}{Bulk}  & 
       \multicolumn{2}{|c|}{Surface}  & 
       \multicolumn{2}{|c|}{Leakage} &
       \multicolumn{2}{|c|}{Improve} 
       & Leakage \\ 
          & all               & \multicolumn{1}{|c|}{singles}           & 
	    all               & singles           & wo/cuts &
	    w/cuts & Clean & Reject & Goal \\
\hline
All      & $6 \times 10^{7}$ & $1.5 \times 10^{7}$  
         & 600000 & $170000$ & 20000~ & 600~~
	 & $2\times$~ & $100\times$ & $<1.0$ \\ \hline
$\gamma$ & $6 \times 10^{7}$ & $1.5 \times 10^{7}$ 
         & 350000 & $100000$ & 2500~ & 70~~
	 & $2\times$~ & $100\times$ & $<0.5$ \\ \hline
$\beta$  & $ 7 \times 10^{5}$ &  $2\times 10^{5}$                     
         & 250000  & 70000 & 17000~ & 500~~
	 & $10\times$~ & $100\times$ & $<0.5$ \\ \hline
	  
\hline
\end{tabular}
\label{twr_rates}
\end{table}

Rejection of photon-related backgrounds in the \CDMSII\ experiment has
been measured 
using high-energy photon sources.
Rejection of photon and electron backgrounds has been simulated using
GEANT4, including
tracking of low-energy electrons created by photon interactions and
inferred depth-dependent 
ionization yield 
based on calibrations with an electron source~\cite{mandic}.
The simulations and measurements 
%agree 
both indicate
that of all photon-related events 
(15-45~keV), 0.3\% are single-scatter surface events 
with some ionization-yield suppression, and 1.8\% of those (or
0.005\% of all photon-related events) suffer enough ionization-yield
suppression to be misidentified as nuclear recoils (see
Table~\ref{twr_rates}).  
Thus, rejection is 99.995\% efficient 
based on the number of scatters and ionization yield
alone, and is  99.9999\% after applying the phonon-timing cut.

\begin{figure}[tbp]
\centering
\includegraphics[width=2.2in]{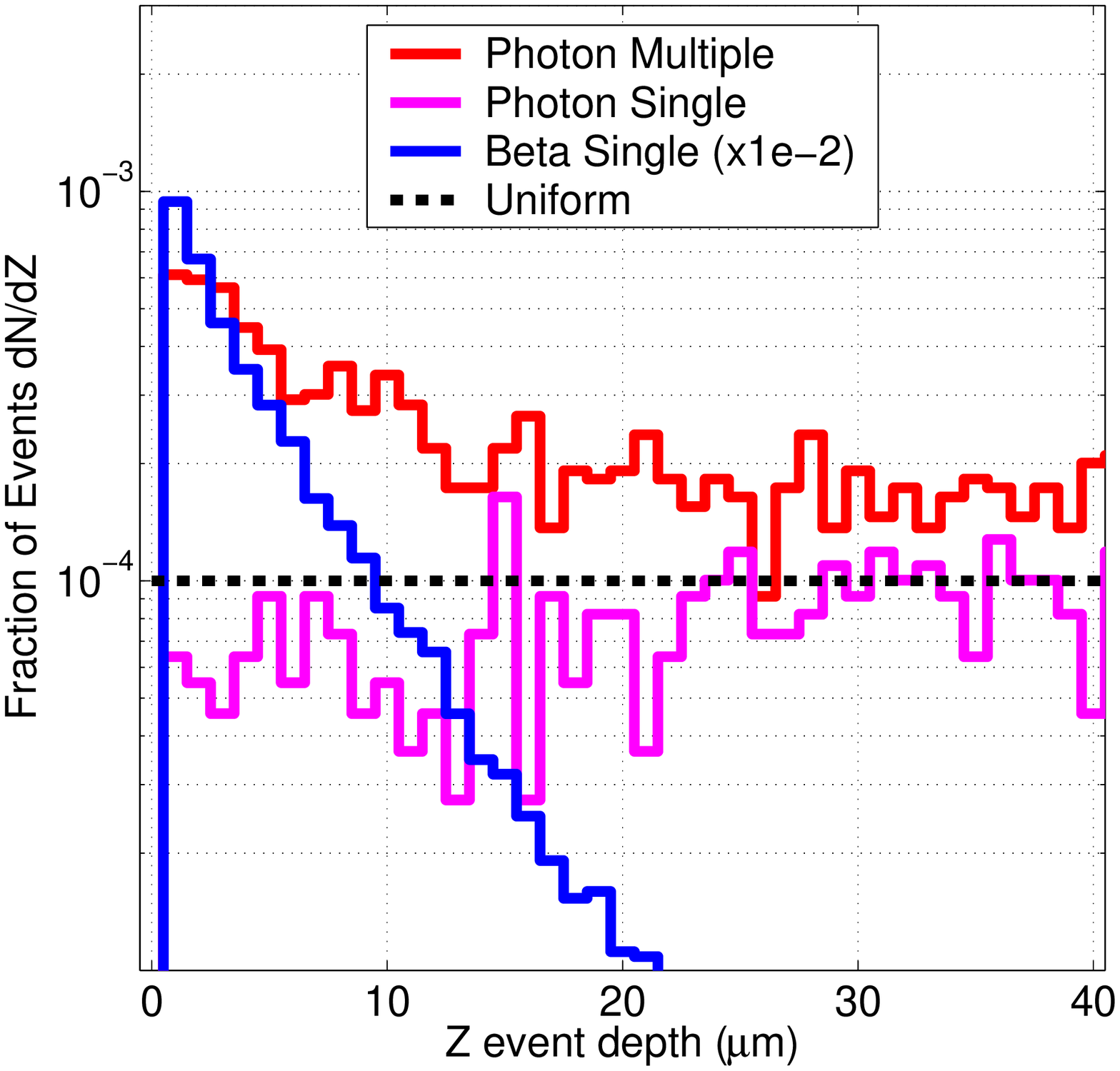}
\includegraphics[width=2.2in]{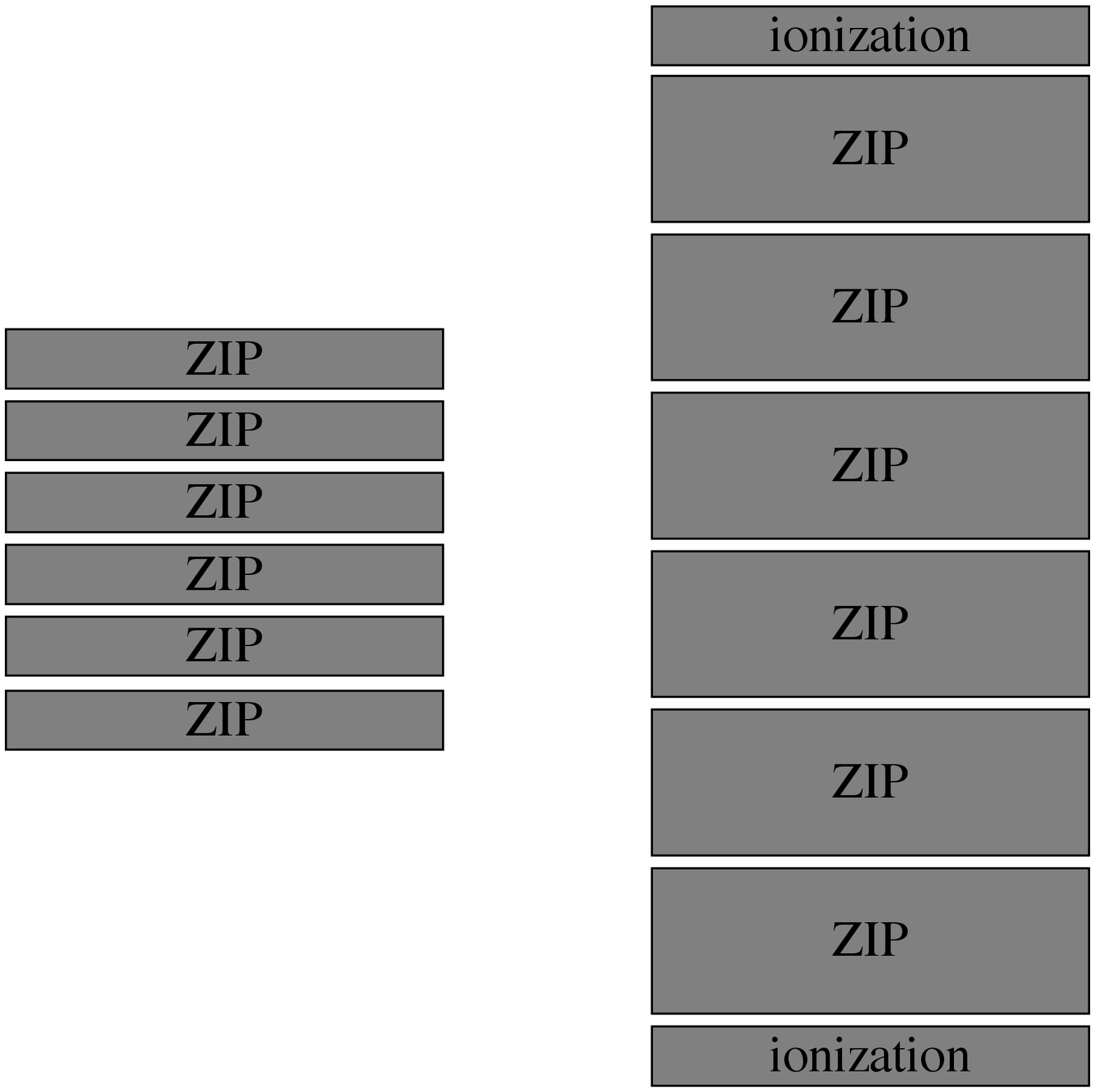}
\caption{\small 
Number of simulated events as a function of depth (left).
Single-scatter photon-induced events (lower grey) are uniformly distributed in
depth, and multiple-scatter events (upper grey) are biased toward the surface.
Events from a beta emitter (black) show a steep falloff with depth,
making them the most dangerous.
Side views show a stack of cylindrical SuperCDMS ZIP
detectors (right) will be $2.5\times$ thicker than \CDMS\ \ZIPs\ (center), 
with ionization endcap
detectors to reduce or veto electrons incident on the outermost ZIP
surfaces.
}
\label{fig:surface_depth_profile}
\end{figure}

Table~\ref{twr_rates} shows that 
the background of surface singles is about twice as large as the inferred 
contribution from photon-related events, implying that about half the 
surface singles are due to surface radioactive beta contaminants.
As shown in Fig.~\ref{fig:surface_depth_profile},
surface betas cause very shallow interactions 
and thus suffer far higher %worse
misidentification than photon-induced events,
resulting in background leakage $\sim7\times$ worse than that due to
photons.

Background-reduction efforts will therefore center on reducing the surface
beta contamination by a factor of 10 by identifying the contaminants
and changing fabrication procedures to prevent their introduction.
Table~\ref{tab:screen} lists screening methods for 
79 beta-emitting and electron-capture isotopes.
Inductively-coupled-plasma mass spectrometry (ICP-MS)
%, with a sensitivity as good as 1~ppt, 
would provide the quickest screening
method for isotopes for which its sensitivity 
(typically 1\,ppb to 1\,ppt)
is good enough.
A dozen isotopes including $^{210}$Pb, a crucial background candidate, 
can be detected by their alpha emissions. 
%of $\alpha$'s.  
An additional 25 isotopes may be detected by low-level $\gamma$-counting.

\begin{table}[tbp]
\caption{Detection schemes for all long-lived beta-emitting isotopes.
Isotopes in boldface may be detected by ICP-MS with sensitivity
between 1\,ppb and 1\,ppt.}
\centerline{\begin{tabular}
{|c|c|}\hline
Method & Applicable Isotopes \\ \hline
ICP-MS  &     $^{40}$K  $^{48}$Ca  $^{50}$V   $^{87}$Rb $^{92}$Nb
              $^{98}$Tc $^{113}$Cd $^{115}$In $^{123}$Te \\
(1~ppb)  &    $^{138}$La $^{176}$Lu $^{182}$Hf $^{232}$Th 
              $^{235}$U $^{238}$U $^{236}$Np $^{250}$Cm \\ 
               \hline
ICP-MS &       $^{10}$Be $^{36}$Cl $^{60}$Fe $^{79}$Se $^{93}$Zr
               $^{94}$Nb $^{97}$Tc $^{99}$Tc $^{107}$Pd $^{126}$Sn
	       \\
(1~ppt) &      $^{129}$I $^{135}$Cs $^{137}$La $^{154}$Eu
               $^{158}$Tb $^{166m}$Ho $^{208}$Bi $^{208}$Po
               $^{209}$Po $^{252}$Es \\ \hline 
$\gamma$ &     $^{40}$K    $^{50}$V   $^{60}$Fe $^{60}$Co $^{93}$Zr
               $^{92}$Nb   $^{94}$Nb  $^{93}$Mo $^{98}$Tc $^{99}$Tc
               $^{101}$Rh  $^{101m}$Rh \\
	   &   $^{102m}$Rh $^{109}$Cd $^{121m}$Sn
               $^{126}$Sn  $^{125}$Sb $^{129}$I $^{134}$Cs $^{137}$Cs
               $^{133}$Ba  $^{138}$La \\
	   &   $^{145}$Pm  $^{146}$Pm $^{150}$Eu
               $^{152}$Eu  $^{154}$Eu $^{155}$Eu $^{157}$Tb $^{158}$Tb
               $^{166m}$Ho $^{173}$Lu \\
	   &   $^{174}$Lu  $^{176}$Lu $^{172}$Hf
               $^{179}$Ta  $^{207}$Bi $^{208}$Bi $^{232}$Th $^{235}$U
               $^{238}$U   $^{236}$Np $^{241}$Pu  
	       \\ \hline
$\alpha$ &     $^{210}$Pb $^{208}$Po $^{209}$Po $^{228}$Ra $^{227}$Ac
               $^{232}$Th \\ 
	 &      
	       $^{235}$U $^{238}$U $^{236}$Np $^{241}$Pu
              $^{250}$Cm $^{252}$Es 
	      \\ \hline
$\beta$ only &  
               $^{3}$H $^{14}$C $^{32}$Si $^{63}$Ni $^{90}$Sr
               $^{106}$Ru $^{113m}$Cd $^{147}$Pm $^{151}$Sm
	       $^{171}$Tm 
               $^{194}$Os\\
	     & $^{204}$Tl 
	      {\bf $^{10}$Be $^{36}$Cl $^{79}$Se $^{97}$Tc
               $^{107}$Pd $^{135}$Cs $^{137}$La $^{154}$Eu
               $^{209}$Po }
	       \\ \hline
\end{tabular}}
\label{tab:screen}
\end{table}

Finally, there are 12 to 21 isotopes, depending on ICP-MS sensitivity,
that cannot be screened in any manner except by their emission of beta
electrons.  To detect these isotopes, we will develop a chamber
capable of directly detecting betas emitted from surfaces.  This
chamber will also serve as a very low-background $\alpha$ screener.

Further reductions in surface backgrounds will be achieved
by decreasing the exposed surface area per detector mass, as shown 
in Fig~\ref{fig:surface_depth_profile}.
The major thrust in detector development
will be
in scaling up the detector thickness from 1~cm to 2.5~cm. 
The thicker detectors
will have a $2.5\times$ smaller surface-to-volume ratio, 
thereby decreasing the background surface events per WIMP
interaction by the same factor regardless of source. 
This increase
requires new fabrication equipment and modifications to some of our
present equipment.
A larger voltage would be applied across these
thicker detectors during their operation,
resulting in only a slightly smaller drift field
that would not significantly reduce the ionization yield for surface 
events.

Ge ionization detectors will act as
an active veto shield around, above, and below the \ZIP\ detectors.
Ionization detectors cost $\sim$5 times
%about a factor of five 
less to fabricate and test than
the \ZIPs. 
These veto detectors may reduce the 
contamination adjacent to
the \ZIPs;
the \CDMSI\ experiment indicated that the Ge
detector material itself has the cleanest surfaces within the detector
housings. 
Moreover, these detectors would reject the otherwise outermost \ZIP\
detectors' dominant background of
single-detector surface events caused by photons ejecting electrons
from adjacent passive materials. 

\section{Detector Performance Improvements}
\label{sect:rejection} 

The complementary method to reduce surface-event backgrounds in SuperCDMS
is to improve the detector rejection of surface events.
Improved analysis already is showing significant advances 
and should increase rejection by at least an order of magnitude.
Further improvements can be achieved 
by optimization of
both the charge collection and the athermal phonon sensors.

First, it is likely that we can improve the charge collection for
surface events by optimizing the deposition of the amorphous-silicon
layer used to prevent 
back-diffusion of carriers to the ``wrong'' electrode~\cite{shutt}.
Older \CDMS\ detectors with a different amorphous-silicon layer
had a higher ionization yield
for surface events than current detectors,
resulting in $>95$\% rejection of surface betas based on ionization 
yield alone\cite{prd19}.
Returning to and possibly improving upon the old recipe should
increase
the blocking effectiveness of the
electrodes against surface-event charge back-diffusion. 

A longer-term goal is to enhance the information in
the athermal phonon signals.  
Reading the phonon signal from the detector substrate faster
should improve the pulse discrimination.
The present \ZIP\ detectors'
phonon sensors  cover only %effectively 
20\% of one surface of the detector
substrate.
New phonon-readout schemes described in Sect.~\ref{sect:build}
would allow increased surface-area coverage.

The most dramatic improvement in phonon read-out
would occur if {\it both} sides of the disc-like detectors were
instrumented with phonon sensors.  This double-sided readout 
should improve the phonon timing information by detecting
the leading phonons on both faces of the detector. This design will
symmetrize the detector response and allow a direct determination of
the three-dimensional position of each event.  The expected timing
resolution of 0.5~$\mu$s at 10~keV would yield a position resolution of 1.5~mm
in three dimensions. 

One method of implementing this scheme 
keeps the same drift-field configuration used in the existing \ZIP\
detectors. The problem then is to provide a voltage bias of up to 1~$\mu$V for
the phonon sensors
without adding more than a few picofarads to the capacitance of
the ionization electrode, 
thereby degrading the ionization resolution.
The best option is a high-frequency (20~MHz) AC power supply
feeding a small transformer.  A diode and LC filter on the secondary
produce the appropriate bias.  The voltage applied could be
regulated at low bandwidth by monitoring the current through the
readout SQUIDs, while maintaining at high bandwidth the pure voltage
bias needed for the phonon sensors.  

An alternative scheme~\cite{luke,cabreraizip} 
is to interleave the ionization electrodes and
the phonon collectors on both detector faces. 
%As shown schematically
%in Fig.~\ref{fig:Interleaved}, o
On one face the ionization electrodes
are at positive voltage and connected to one
amplifier, while on the other face they are at negative voltage 
and connected to a second amplifier.  The phonon sensors
remain at ground on both faces. In this arrangement,
the electric field close to the surfaces is roughly parallel to the
detector face and so a surface interaction generates a signal in the
near-side
charge amplifier only. Bulk events experience a field 
perpendicular to the faces, causing the electrons and holes from an event
to generate equal signals in each charge amplifier. 
The interleaved design thus provides an additional means of
discriminating surface events from bulk events.

%\begin{figure}[tbp]
%\centering
%\begin{center}
%\vspace{0.2in}
%\includegraphics[width=3.0in]{Interleaved_Q_P.eps}
%\caption{\small Electric field lines resulting from double-sided
%phonon sensor design (viewed edge-on) 
%with  phonon sensors at ground (small circles) interleaved
%with ionization channels (large circles).
%On the top face, the ionization electrodes
%are at positive voltage,  
%while on the other face they are at negative voltage. 
%There is also a ground ring around the side.  
%Because the electric field close to the surfaces is roughly parallel to the
%detector face, a surface interaction generates a signal in the
%%near-side charge amplifier only.
%Bulk events are seen equally in both charge
%channels.}
%\label{fig:Interleaved}
%\end{center}
%\end{figure}

Since the interleaved
phonon sensors on both sides of the crystal remain at ground for the
ionization measurement, 
the biasing electronics remain
unchanged from the present implementation.  
%In addition, i
If the
surface identification works at the outer edges of the detector, 
no outer guard electrode is needed, and the readout
and electronics 
of the two
ionization channels 
also remain
unchanged from the the present \ZIP\ implementation of \CDMSII.

\section{Improving Manufacturability}
\label{sect:build}

At the \CDMSII\ detector production rate of about one detector per
month, or 3~kg~Ge per year, deploying a ton of detectors would be
impossible.  Fortunately, nearly all the time associated with detector
production is spent on a lengthy program of testing and repairing,
which could be rendered unnecessary by
improvements to the fabrication process.
These improvements would 
allow orders of 
magnitude increases in detector production.  Furthermore, many of
these fabrication 
improvements also allow easier building of the experimental infrastracture
and decrease the cryostat's material and heatload. 

Several straightforward improvements are planned to increase
fabrication yield.  Changing to a whole-field mask for first-layer
exposures, optimizing etching recipes, 
and switching from 1\,$\mu$m to 2\,$\mu$m features 
should greatly reduce photolithography errors and the resulting
need for testing and repair.  These changes should increase the
fabrication rate by a factor of 5.

Further increases require reducing or eliminating the need for 
cryogenic testing of the W thin-film critical temperature.
The addition of a mechanical planetary system for W film depositions
may result in reproducible, uniform critical temperatures.
The sensitivity of W to processing conditions has, in addition,
motivated work to develop alternative films such as Al(Mn) that
may be more reproducibly processed~\cite{AlMn}. 
Finally, studying W film properties
(crystalline phase, resistivity, film thickness, crystallite size and
sputtering conditions) 
may lead to the establishment of room-temperature %
diagnostic 
tests.

\subsection{Lower-Inductance  SQUID Readout}

In \CDMSII\ the transition-edge sensors (\TESs)
of each phonon sensor are in series with
the input coil to a SQUID array of 100 elements.  A voltage is applied
across the combined \TES/input coil pair, and the SQUID array output
measures the current through the \TES. 
In order to have sensitivity to resolve the \TES\ current of phonon events,
the SQUID array must have
many turns on the input coil for flux amplification.  
However, 
the ratio of the
input inductance of the SQUID array to the operating dynamic resistance of the
\TES, $L_{\mathrm{in}}$(array)/$R_{\mathrm{dyn}}$(\TES),
sets a limit on bandwidth.
%In order f
For the phonon sensor
readout to have sufficient bandwidth ($\sim$100~kHz) and to be stable
from electrothermal oscillations,
%can cut this phrase:
the dynamic resistance of the phonon sensor
$R_{\mathrm{dyn}} > 0.1 \Omega$.

A number of the proposed advances of our phonon
sensor technology described above would benefit from a decrease of
$L_{\mathrm{in}}$. 
These include the increase of the coverage of the phonon sensors, which
would require connecting more \TESs\ in parallel; the widening of the W
\TES\ to ease the fabrication; 
and the use of Al(Mn), which has intrinsically lower
resistivity. Moreover, the
phonon rejection would likely benefit from the increased speed.

Two approaches are possible. 
One 
%approach
is to decrease the
apparent input inductance through feedback.  
%One version of this method replaces 
Replacing our current magnetic
feedback through a feedback coil with a resistive feedback to the input
of the SQUID
would decrease the effective inductance by the gain of the feedback loop. 
This scheme requires no modification of the warm electronics and only a minor
rewiring on the cold electronics stage.  

A second approach is to use a single SQUID front end, so that the
input inductance is physically smaller.  This scheme requires a
two-stage SQUID configuration, such as has been operated for many
years~\cite{irwin_twostage}.  
A voltage-biased single SQUID is 
the initial sensor of \TES\ current.  The current from the single
SQUID is amplified by a second-stage SQUID array 
similar to the single-stage array used in \CDMSII.
A two-stage system has greater current sensitivity at the
first-stage input and hence greater signal to noise.
%, providing
%additional design flexibility for TES biasing schemes.
%The cold electronics hardware and the room temperature electronics
%would share similarities to that discussed above for the SQUID charge
%readout scheme.
The two-stage system also imposes less stringent
design and fabrication 
requirements on the SQUID array
%, resulting in
.  With a
first-stage preamplifier SQUID, less gain is required in the SQUID
array; the fewer turns on the input coil result in 
greater reliability in the SQUID fabrication and operation.

\subsection{SQUID Ionization Readout} 
\label{subsect:charge}

Schemes using SQUID amplifiers to read out the ionization channels
would eliminate the need for tensioned signal wires
going inside vacuum coax from 50~mK to 4~K, thus allowing 
the bulky, massive mounting hardware to be made much lighter and more flexible.
Using SQUIDs would also decrease radically the power dissipated, 
an important
consideration given the number of channels needed for SuperCDMS.

To provide enough
sensitivity to measure the ionization current with SQUIDs, 
a high turn-ratio superconducting
transformer (1:2000) would be coupled to a two-stage SQUID. 
The challenges of controlling stray capacitance and
internal resonance for such a large transformer appear manageable,
and
performance 
at least as good as that of \CDMS\
appears achievable~\cite{irwincharge}.  
This approach uses the same type of amplifier as the phonon
sensors, thus simplifying the overall electronics
systems.

\section{Conclusions}

Modest improvements in the level and/or discrimination of backgrounds are needed
to keep backgrounds negligible during the three phases of SuperCDMS.
By developing production designs that require only modest testing, 
detector production rates may be improved sufficiently
to allow an exposure of 500~ton~d within a reasonable time and budget.
Overall, the improvement estimates described above are conservative.
Previous development
efforts have shown that some areas prove easier and provide larger factors while
others prove more difficult. The conservative estimates together with
the broad approach reduce the risk and give us confidence that we will
succeed, providing the surest way to probe to WIMP-nucleon cross
sections of 10$^{-46}$\,cm$^2$.

%\begin{table}
%\centering
%\begin{tabular}{lll}
%\hline\noalign{\smallskip}
%\end{tabular}
%\end{table}
%
%
%
%
% BibTeX users please use
%\bibliographystyle{plain}
%\bibliography{dark2004schnee}
%
% Non-BibTeX users please follow the syntax
% the syntax of "referenc.tex" for your own citations
%\input{referenc}
\gdef\journal#1, #2, #3, #4#5#6#7{ #1~{\textbf {#2}}, #3 (#4#5#6#7)} 

\def\arp{\journal Ann.\ Rev.\ Nucl.\ Part.\ Sci., }
\def\apl{\journal Appl.\ Phys.\ Lett., }
\def\apj{\journal Astrophys.\ J., }
\def\app{\journal Astropart.\ Phys., }
\def\baas{\journal Bull.\ Am.\ Astron.\ Soc., }
\def\ejpc{\journal Eur.\ J.\ Phys.\ C., }
\def\jltp{\journal J.\ Low\ Temp.\ Phys., }
\def\jhep{\journal J.\ High Energy Phys., }
\def\nature{\journal Nature, }
\def\nc{\journal Nuovo Cimento, }
\def\nima{\journal Nucl.\ Instr.\ Meth.\ A, }
\def\np{\journal Nucl.\ Phys., }
\def\npps{\journal Nucl.\ Phys.\ (Proc.\ Suppl.), }
\def\pl{\journal Phys.\ Lett., }
\def\prep{\journal Phys.\ Rep., }
\def\pr{\journal Phys.\ Rev., }
\def\prc{\journal Phys.\ Rev.\ C, }
\def\prd{\journal Phys.\ Rev.\ D, }
\def\prl{\journal Phys.\ Rev.\ Lett., }
\def\rnc{\journal Riv.\ Nuovo\ Cim., }
\def\rsi{\journal Rev. Sci. Instr., }
\def\rpp{\journal Rep.\ Prog.\ Phys., }
\def\sjnp{\journal Sov.\ J.\ Nucl.\ Phys., }
\def\solarphys{\journal Solar Phys., }
\def\jetp{\journal J.\ Exp.\ Theor.\ Phys., }

%%%%%%%%%%%%%%%%%%%%%%%%%%%%%%%%%%%%%%%%%%%%%%%%%%%%%%%%%%%%%%%%%%%%%%  }

%%%%%%%%%%%%%%%%%%%%%%%%%%%%%%%%%%%%%%%%%%%%%%%%%%%%%%%%%%%%%%%%%%%%%%

\printindex
\end{document}